\documentclass[aps,prl,showpacs,notitlepage,nofootinbib,superscriptaddress,floatfix,showkeys,twocolumn]{revtex4-1}

\usepackage{blindtext}
\usepackage{hyperref}
\usepackage{amsmath,amssymb}
\usepackage{float}
\usepackage{microtype}
\usepackage{graphicx}
\usepackage{bm}
\usepackage{latexsym}
\usepackage{epsfig}
\usepackage{psfrag}
\usepackage{color}
\usepackage[dvipsnames]{xcolor}
\usepackage{subfigure}
\usepackage[section]{placeins}
\def\e1i{\epsilon_{1\mathrm{i}}}

\allowdisplaybreaks[1]

\newcommand{\mnras}{Mon. Not. Roy. Astron. Soc.\,}

\begin{document}

\title{Light scalar explanation for 18~TeV GRB 221009A}
\author{Shyam Balaji}
\email{sbalaji@lpthe.jussieu.fr}
\affiliation{Laboratoire de Physique Th\'{e}orique et Hautes Energies (LPTHE),\\
UMR 7589 CNRS \& Sorbonne Universit\'{e}, 4 Place Jussieu, F-75252, Paris, France}
\affiliation{Institut d’Astrophysique de Paris, UMR 7095 CNRS \& Sorbonne Universit\'{e}, 98 bis boulevard Arago, F-75014 Paris, France}
\author{Maura~E.~Ramirez-Quezada}
\email{me.quezada@hep-th.phys.s.u-tokyo.ac.jps}
\affiliation{Department of Physics, University of Tokyo, Bunkyo-ku, Tokyo
 113--0033, Japan}
\affiliation{Dual CP Institute of High Energy Physics, C.P. 28045, Colima, M\'exico}
\author{Joseph Silk}
\email{silk@iap.fr}
\affiliation{Institut d’Astrophysique de Paris, UMR 7095 CNRS \& Sorbonne Universit\'{e}, 98 bis boulevard Arago, F-75014 Paris, France}
\author{Yongchao Zhang}
\email{zhangyongchao@seu.edu.cn}
\affiliation{School of Physics, Southeast University, Nanjing 211189, China}
\begin{abstract}
The reecent astrophysical transient Swift J1913.1+1946 may be  associated with the $\gamma$-ray burst GRB 221009A. The redshift of this event is $z\simeq 0.151$. Very high-energy $\gamma$-rays (up to 18 TeV) followed the transient and were observed by LHAASO; additionally, Carpet-2 detected a photon-like air shower of 251 TeV. Photons of such high energy are expected to readily annihilate with the diffuse extragalactic background light (EBL) before reaching Earth. If the $\gamma$-ray identification and redshift measurements are correct, new physics could be necessary to explain these measurements. This letter provides the first CP-even scalar explanation of the most energetic 18 TeV
event reported by LHAASO. In this minimal scenario, the light scalar singlet $S$ mixes with the Standard Model (SM) Higgs boson $h$. The highly boosted $S$ particles are produced in the GRB and then undergo the radiative decay di-photon $S\rightarrow \gamma\gamma$ while propagating to Earth. The resulting photons may thus be produced at a remote region without being nullified by the EBL. Hence, the usual exponential reduction of $\gamma$-rays is lifted due to an attenuation that is inverse in the optical depth, which becomes much larger due to the scalar carriers.

\end{abstract}
\maketitle

\section{Introduction}

Gamma-ray burst GRB 221009 A events over energies from $500\,\rm GeV$ extending up to $18\,\rm TeV$ have been recently detected. The events are reported to have occurred at a redshift of $z\approx 0.15$ corresponding to a distance of $d=645\,\rm Mpc$~\cite{2022GCN.32648....1D}. The initial detection was first reported by the {\it Swift}-Burst Alert Telescope~\cite{GRB20221009A}, the Fermi satellite~\cite{2022GCN.32636....1V} and CARPET-2~\cite{2022ATel15669....1D}. The number of photons detected by LHAASO \cite{2022GCN.32677....1H} in an interval of $2000$ s was $\mathcal{O}(5000)$ events.  However, observing such energetic photons is extremely unlikely since the flux is expected to be rapidly attenuated when propagating and interacting with extragalactic particles\cite{PhysRevLett.16.252,osti_4836265,Fazio:1970pr, Balaji:2022wqn}. The unlikeliness for photons in the upper band of the $500\,{\rm GeV} $-$18\,\rm TeV$ range to survive in this manner while propagating to the Earth motivated several studies beyond the Standard Model (SM) as possible solutions. These 
beyond-the-SM models include explanations from sterile neutrinos~\cite{Smirnov:2022suv,Brdar:2022rhc,Huang:2022udc,Guo:2023bpo}, axion-like particles\cite{Baktash:2022gnf,Galanti:2022pbg,Lin:2022ocj,Troitsky:2022xso,Nakagawa:2022wwm} and axion-photon conversion scenarios \cite{Zhang:2022zbm}. Even SM explanations have recently been discussed, where external inverse-Compton interactions with the cosmological radiation fields and proton synchrotron emissions could generate the 
ultra-energetic $\gamma$-rays~\cite{Zhang:2022lff,AlvesBatista:2022kpg,Gonzalez:2022opy,Sahu:2022gvx}.

In this work, we consider another possible scenario that can explain the high energy $\gamma$-ray emission. Here, we consider a generic scalar $S$, corresponding to  the inclusion of a gauge singlet to the SM scalar sector. This new 
field couples to the SM particles through mixing with the SM Higgs field $h$, and the interaction is  parameterised by the  mixing parameter $\sin\theta$ with angle $\theta$. The singlet extension of the SM scalar sector is one of the simplest underlying models, and has a host of motivations including aiding in stabilizing the SM vacuum~\cite{Gonderinger:2009jp, Gonderinger:2012rd, Lebedev:2012zw, Elias-Miro:2012eoi, Khan:2014kba, Falkowski:2015iwa, Ghorbani:2021rgs}, addressing the hierarchy problem in relaxion models~\cite{Graham:2015cka, Flacke:2016szy, Frugiuele:2018coc, Banerjee:2020kww, Brax:2021rwk}, generating the observed baryon asymmetry in the Universe via baryogenesis~\cite{Espinosa:1993bs, Choi:1993cv, Ham:2004cf,Profumo:2007wc,Espinosa:2011ax, Barger:2011vm,  Profumo:2014opa,Curtin:2014jma, Kotwal:2016tex,Chen:2017qcz,Balaji:2020yrx}, addressing the cosmological constant problem through radiative breaking of classical scale invariance~\cite{Foot:2011et, Heikinheimo:2013fta, Wang:2015cda}, and mediating the interactions between dark matter (DM) and the SM sector~\cite{Pospelov:2007mp, Baek:2011aa, Baek:2012uj, Baek:2012se, Schmidt-Hoberg:2013hba, Krnjaic:2015mbs, Beniwal:2015sdl,Balaji:2018qyo}. 

To accommodate the aforementioned high energy photon observations, the $S$ particles corresponding to this new field are produced in large numbers during the GRB through hadronic scattering and then undergo the radiative decay $S\rightarrow \gamma\gamma$ while propagating to Earth. The resulting decay photons may therefore be produced at a region with less interstellar medium and avoid being nullified. Hence, the $S$ particles provide an effective means for the survival of the photons to Earth. More explicitly, this corresponds to the usual exponential reduction of $\gamma$-rays being lifted due to an attenuation that is inverse in the optical depth.

In this work, we will use the 18 TeV GRB $\gamma$-ray flux observation and the number of the detected events in the $2000$ second observation window to set new astrophysical bounds. These limits complement the existing bounds derived from stellar cooling and luminosity measurements of the Sun, white dwarfs, red giants and supernovae~\cite{Balaji:2022noj,Dev:2020jkh,Dev:2021kje}.

The outline of this paper is as follows. In Sec.~\ref{sec:model} we introduce the simplified extension to the SM scalar sector and show the decay rates of the scalar $S$ into various states. In Sec.~\ref{sec:flux} we compute the flux of $\gamma$-rays produced by the S-decay. Finally, we summarise and discuss the GRB bounds on the scalar mass and mixing in Sec.~\ref{sec:results} and conclude in Sec.~\ref{sec:conclusion}.

\section{Scalar decay}
\label{sec:model}
This section reviews the decay rates of the CP-even scalar $S$ into $\gamma$-rays. Therefore, we first introduce the scalar interactions with the SM particles. At tree-level, the $S$ interacts with leptons and pions via mixing with the SM Higgs. The interaction Lagrangian for these specific interactions is given by
\begin{equation}
    \mathcal{L}=-\sin\theta S \left[ A_\pi (\pi^0\pi^0+\pi^+\pi^-)+\frac{m_\ell }{v}\Bar{\ell}\ell \right] \,,
\end{equation}
where $A_\pi$ is the effective coupling of $S$ to pions~\cite{Voloshin:1985tc,Donoghue:1990xh}. $\sin\theta$ is the $S$-mixing with the SM Higgs $h$ and $\ell$ denotes leptons.\footnote{Technically $\ell$ represents electrons, muons and $\tau$ leptons, however in this work, we only consider electrons but do not consider scalars heavy enough to decay into muons and $\tau$ leptons.} 
Hence, the scalar can  decay into photons at one-loop level as well as electrons, muons and pions at tree-level from mixing with the Higgs boson. 

For $S$ scalars, with masses below twice the electron mass ($m_S<2m_e$), the dominant channel is the decay to two photons ($S\to \gamma\gamma$). The decay rate for this process is given by~\cite{Dev:2017dui},
\begin{equation}
    \Gamma_{S\gamma\gamma}= \frac{121}{9} \frac{\alpha^2 m_S^3 \sin^2\theta}{512 \pi^3 v^2} \,,
\end{equation}
where $\alpha=e^2/4\pi$ is the  fine structure-constant and $v=246\,\rm GeV$ is the Higgs vacuum expectation value.

When the  $S$ scalar  masses are above twice the lepton mass threshold ($m_S\gtrsim 2m_\ell$), the scalar can decay into two oppositely charged leptons respectively. In this case, the decay rate  ($S\to \ell^+\ell^-$)  is given by~\cite{Dev:2020eam},
\begin{align}
    \Gamma_{S\ell^+\ell^-}&=\frac{m_Sm_\ell^2 \sin^2\theta}{8\pi v^2}\left(1-\frac{4m_\ell^2}{m_S^2}\right)^{3/2}
\end{align}
where $m_{\ell}$ is either the electron or muon mass depending on the decay channel of interest. Here, we will only consider the scalar massive enough to decay into a pair of electrons $e$.

\section{Photon flux}
\label{sec:flux}

We now compute the $\gamma$-ray flux  $\Phi_\gamma$ from the $S$ scalar decay. First, we introduce the $\gamma$-ray flux $\Phi_\gamma^0$ of GRB 221009A, which corresponds to the flux where all photons survive their interactions with the extragalactic medium on their way to the Earth, 
\begin{align}
    \Phi_\gamma^0(E_\gamma) =& 2.1\times 10^{-6} \, {\rm TeV}^{-1} \, {\rm cm}^{-2} \, {\rm s}^{-1} \nonumber \\ & \times \left(\frac{E_\gamma}{\rm TeV}\right)^{-1.87\pm0.04}.
    \label{eq:unattenuatedflux}
    \end{align}
This is the so-called {\it unattenuated flux}. As discussed in Ref.~\cite{Smirnov:2022suv}, this is ascertained by extrapolating the flux measured by Fermi-LAT (GCN 32658) in [0.1,1] GeV to TeV scale~ \cite{Baktash:2022gnf}.

We define the decay-production probabilities that a single particle $S$ decays into two photons, and they reach the Earth. If the $S$ scalar decay occurs at a distance interval of $[x,x+dx]$, the decay-production probability is then given by~\cite{Smirnov:2022suv},
\begin{equation}
   P_{\rm {decay} } =  B_\gamma e^{-x/\lambda_S} \times \frac{dx}{\lambda_S}e^{-(d-x)/\lambda_\gamma}\label{eq:Pdecay}.
\end{equation}
$d$ is the distance at which the flux was produced. 
$B_\gamma$ is the branching ratio for the scalar decay into photons. The second exponential factor is the probability that the produced $\gamma$-rays survive their trajectory and is given in terms of the photon absorption length (or optical depth) $\tau=d/\lambda_\gamma$ where $\lambda_\gamma$ is the mean-free-path of the photon. 
In this expression, the  $\lambda_S$ is the  mean free path (MFP) of the $S$ scalar in terms of the $S$ decay rate $\Gamma_S$ and energy of the scalar $E_S$ in the Earth's rest frame
\begin{equation}
    \lambda_S=\frac{E_S}{m_S \Gamma_S}.
\end{equation}

To compute the $\gamma$-ray flux produced from the $S$ scalar decay, we integrate the expression given in Eq.~\eqref{eq:Pdecay} over $x$ and multiply by the incoming $S$ scalar flux at Earth $\Phi_S$ from the GRB source, that would be expected if the scalar did not decay
\begin{equation}
    \Phi_\gamma =  \frac{2\Phi_S B_\gamma}{\lambda_S/\lambda_\gamma-1}\left(e^{-d/\lambda_S}-e^{-d/\lambda_\gamma}\right),
\end{equation}
the factor of 2 is included to account for the  two photons produced in the $S$-decay. Rewriting  this expression in terms of the photon absorption length $\tau(E_\gamma)$ leads to 
\begin{equation}
    \Phi_\gamma = \frac{2\Phi_S B_\gamma}{\tau \lambda_S/d-1}\left(e^{-d/\lambda_S}-e^{-\tau}\right).\label{eq:flux}
\end{equation}

\begin{figure}
    \centering
    \includegraphics[width=0.48\textwidth]{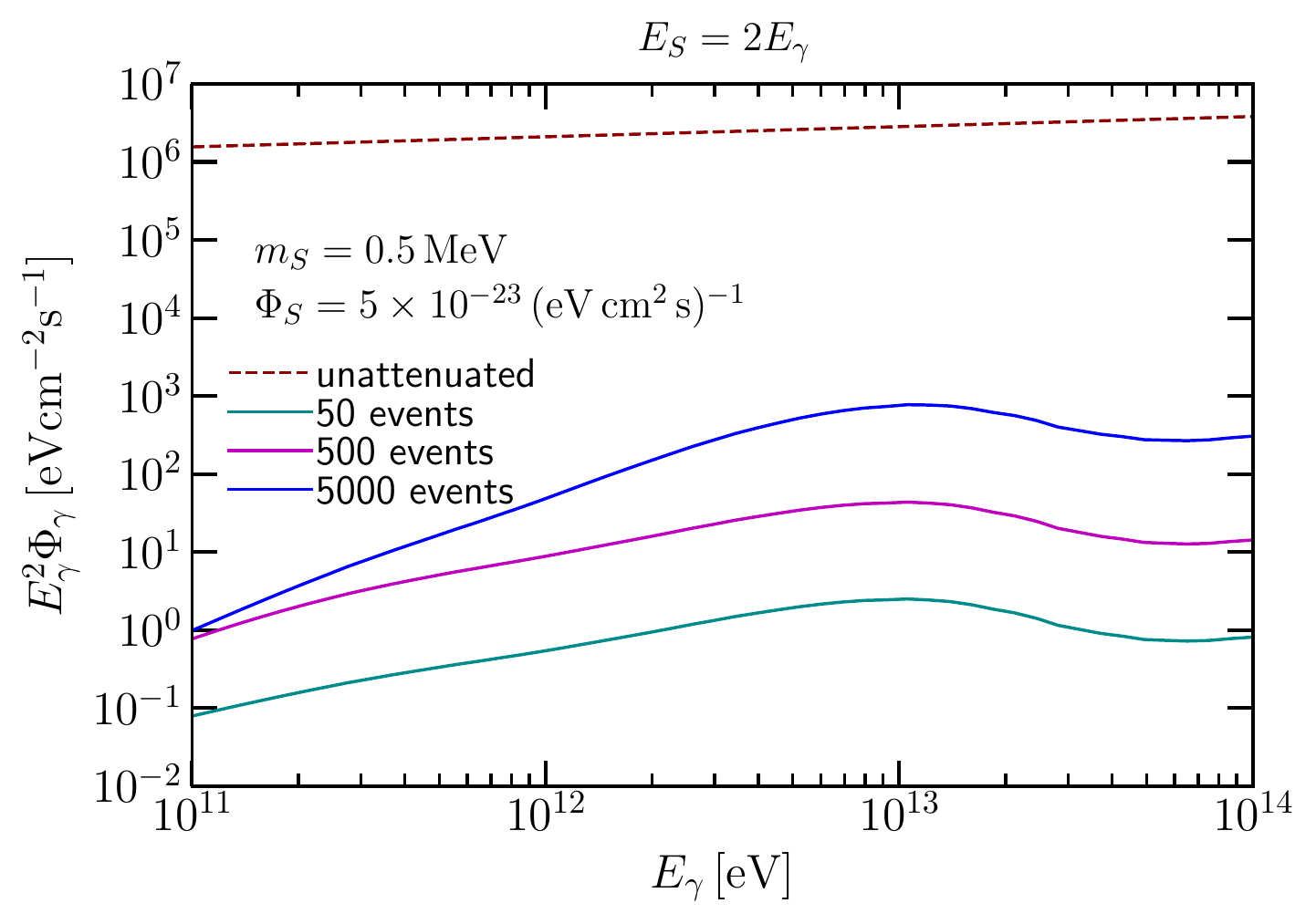}
    \caption{Photon energy flux from GRB 221009A at Earth as a function of photon energy $E_\gamma$ for a scalar flux of  $\Phi_S=5\times10^{-23}\, \rm (eV cm^2 s)^{-1}$ for the cases of 50 events (cyan),  500 events (magenta) and 5000 events (blue). Flux extrapolated from Fermi-LAT (GCN 32658) is $\Phi_\gamma^0$ from Eq.~\eqref{eq:unattenuatedflux} (red-dashed). The energy of the emitted scalars is taken to be $E_S=2\,E_\gamma$ TeV}
    \label{fig:flux}
\end{figure}

We show the $\gamma$-ray flux from the $S$-decay as a function of the photon energy $E_\gamma$ in Fig.~\ref{fig:flux}, the $S$ scalar energy is $E_S=2\,E_\gamma$ with a mass of $m_S=0.5\,\rm MeV$. The mixing has been fixed by requiring three different scenarios, one where the scalar decay leads to 5000 events (blue), 500 events (magenta), and 50 events (cyan). We use the full energy dependence of $\tau(E_\gamma)$ from Ref.~\cite{2011MNRAS.410.2556D} for the flux calculation. 

Since there are large uncertainties in the hadronic environment and energy profile of particles at the source, we do not have exact knowledge about the number of high-energy
 nucleons that can be produced or preexist in that region. However, we know that the dominant mode for light scalar production is nucleon-nucleon bremsstrahlung via pion exchange. 
\cite{Krnjaic:2015mbs,Ishizuka:1989ts,Arndt:2002yg,Diener:2013xpa,Tu:2017dhl,Lee:2018lcj,Dev:2020eam,Dev:2021kje}. Possible sources for very high energy scalars in high nucleon density environments could be via the interaction of AGN jets with obstacles, such as supernova remnants \cite{Torres-Alba:2019bza}. The collision of the supernova ejecta with the high-energy jets could lead to the formation of an interaction region, in which particles can be accelerated and produce high-energy scalar emission \cite{Vieyro:2019smo}. Scalars can also be produced by heavy meson decays. With present information about the GRB, the exact fraction of energy that can be extracted from the GRB in exotic states is unknown. However we can make some conservative estimates and adjust them in them in the event of more refined measurements in the future. The  amount of energy emitted in a  supernova explosion is typically about $10^{44}\,\textrm{J} \approx 6.2 \times 10^{62}$~eV. In general, almost all of this energy is carried away by neutrinos. In the case of the GRB of interest here, the energy output is estimated at $2\times 10^{54}\,\textrm{erg}=1.2\times 10^{66}$~eV \cite{Cheung:2022luv}. Assuming that almost all the energy is lost and emitted as neutrinos, and using 0.3\% of this energy as an upper bound for scalar emission with a rectangular energy distribution in the range $[1,36]$ TeV corresponding to an average energy of $\overline{E}_S=18.5$ TeV \footnote{Note that the rectangular spectrum in the energy of the scalar flux at the GRB source between $[1,36]$ TeV is chosen because this results in photon energies of $E_\gamma \simeq E_S/2$ which lies in the $[0.5,18]$ TeV range LHAASO observations window. In the absence of more information about the GRB source energy profile, this is the simplest possible choice. }, we get a scalar flux (assuming no decays) of $\Phi_S \simeq 5\times 10^{-23}\,\textrm{eV}^{-1} \, \textrm{cm}^{-2}\, \textrm{s}^{-1}$ at Earth over the 2000~s duration. It should be noted that this is very conservative since when considering supernova bounds, typically, the flux of the new particle is bounded by 10\% of the neutrino luminosity, see, e.g. Refs.~\cite{Balaji:2022noj,Arndt:2002yg,Lee:2018lcj}. It should be noted that $E_S=2\,E_\gamma$ is fixed simply due to kinematics of two-body decays. Since the max energy for the GRB photons is observed at 18~TeV, the energy of the parent scalar has to be at least twice this energy. Other, more exotic beyond the SM scenarios can also be invoked to generate the scalar flux from the source, but we do not explore them in this work.

Finally, we can observe in Fig.~\ref{fig:flux} that the flux of $\gamma$-rays from the $S$ scalar decay increases at high energies, approaching the unattenuated $\gamma$-ray flux. This indicates that the photons can survive interactions with the extragalactic background. 

\section{Bounds on the scalar mass-mixing parameter space}
\label{sec:results}

The flux in the previous section provides us with a means to set bounds on the scalar mass-mixing parameter space. We require the number of photons produced by the scalar decay to be  $N_\gamma^S =50,\, 500$ and  5000 respectively. Here $N_\gamma^S$ is obtained by integrating Eq.~\eqref{eq:flux} over the photon energy $E_\gamma$ over the range of [$0.5,18$]~TeV as well as integrating over the  time $t$ from 0 s to 2000 s and the detector area of $1\,\rm{km}^2$ \cite{CUI201486,Ma_2022}.

We consider scalars with masses up to 2 MeV. Therefore, these scalars can only decay directly into photons and electrons. It is easy to find and extrapolate  the relationship between $\sin\theta$ and $m_S$ as a function of the  number of events expected in 2000 s. This mixing-mass relationship providing upper bounds is found to be given by the relation

\begin{equation}
    \sin\theta\simeq \left(\frac{m_S^2}{1.57\times10^{-8}\, \rm{eV^2}} \right)^{-1} \sqrt{\frac{N_\gamma^S }{\Phi_S/(\rm{eV \, cm}^2 \, \rm{s})^{-1}}}\label{eq:Apprelation}.
\end{equation}
\begin{figure}
    \centering
    \includegraphics[width=0.48\textwidth]{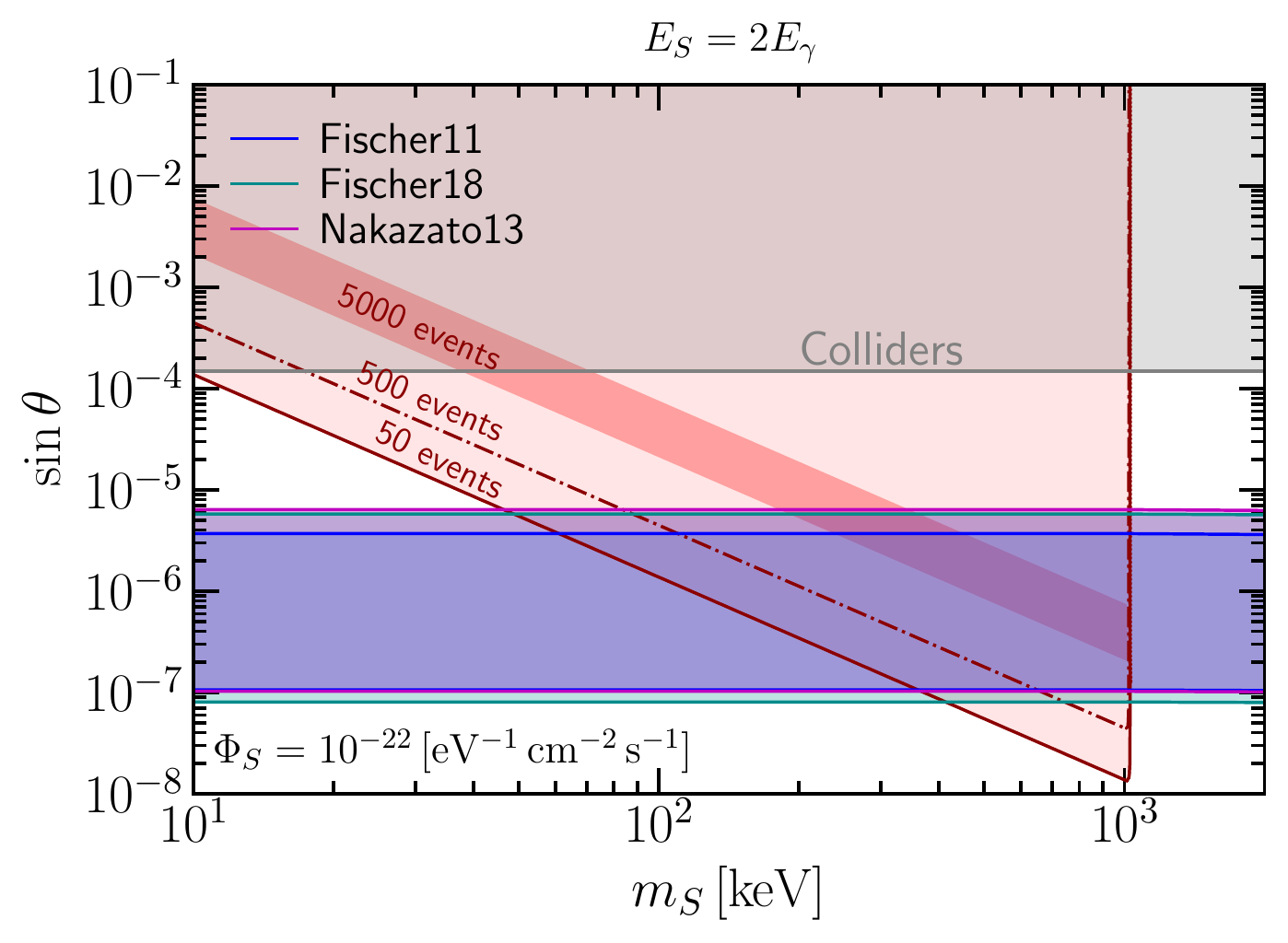}
    \caption{Limits for light scalar decays into photons to explain GRB 221009 A in the scalar mass-mixing ($m_S,\sin\theta$) plane requiring  $N_\gamma^S \geq 50,\, 500$ and  5000 number of events. The shaded regions correspond to supernova exclusions from Ref.~\cite{Balaji:2022noj} from various models such as Fischer 11$M_\odot$ (blue), Fischer 18$M_\odot$ (cyan) and Nakazato 13$M_\odot$ (magenta). Additionally, we add complementary limits from SN1987A and collider searches~\cite{Dev:2017dui,Egana-Ugrinovic:2019wzj,Dev:2019hho}.}
    \label{fig:Limits1}
\end{figure}
This relationship also depends on the scalar flux $\Phi_S$ and assumes an even distribution of $\Phi_S$ over the $E_S$ range $[1,36]$ TeV. 

In Fig.~\ref{fig:Limits1}, we show the upper bounds for light scalar decaying into photons to explain GRB 221009 A in the scalar mixing-mass plane. This was computed numerically using the expression in Eq.~\eqref{eq:flux} considering three benchmarks for the number of events $N_\gamma^S =50,\, 500$ and  5000. We also show the stellar exclusions in Ref.~\cite{Balaji:2022noj} given by the supernova SN1987A using different numerical profiles Fischer 11.8M$_\odot$ (blue), Fischer 18M$_\odot$ (cyan)~\cite{Fischer:2016cyd} and Nakazato 13M$_\odot$ (magenta)~\cite{Nakazato:2012qf}. We also show the combined collider searches (gray)~\cite{Dev:2017dui,Egana-Ugrinovic:2019wzj,Dev:2019hho}.
Note that when there is an order of magnitude difference between the expected number of events, the difference between the bounds is a factor of a few.  For the case of two orders of magnitude difference in $N_\gamma^S$, we find an order of magnitude difference between the limits. This is easily understood from the proportionality between the mixing and the number of events expected scales like $\sin\theta\propto\sqrt{N_\gamma^S}$. Additionally, we check the MFP of the scalar at an average energy of $\overline{E}_S=18.5$ TeV for the three different expected event boundaries shown in Fig.~\ref{fig:Limits1}, these are on the order of $\simeq 10^{30}\,\rm cm$ (50 events), $\simeq 10^{29}\,\rm cm$ (500 events) and $\simeq 4\times10^{26}$-$5\times10^{27}\,\rm cm$ (5000 events) which is comparable to the estimated GRB source distance of $\simeq 10^{27}$~cm.

The bounds obtained from GRB 221009 A  by requiring $N_\gamma^S=50,\, 500$ and $5000$ events complement those given by supernovae~\cite{Balaji:2022noj} and collider exclusions~\cite{Dev:2017dui,Egana-Ugrinovic:2019wzj,Dev:2019hho}. Note that considering regions with $>500$ or $>50$ events, there is no upper bound on mixing. However, there is a small band in Fig.~\ref{fig:Limits1} for $>5000$ events because, in this region, $\lambda_S \simeq d$ and hence the photon flux is locally maximised. 
This can be understood as follows:  At higher mixing, the scalars decay closer to the source, and more of the resulting photons get attenuated by the EBL. 
However, even if the scalars decay very quickly near the source, some minimum number of lower energy photons in the $[0.5,18]$ TeV range will still survive their journey to Earth, albeit with a smaller number of events, as shown. 
The lower bounds on mixing are because the scalars are too long-lived. On average, they may decay well beyond the Earth, reducing the observed photon flux and hence the event yield. 

While supernova bounds excluded parameter space across the whole scalar mass range between mixing of the order of $10^{-7}$-$10^{-5}$, the excluded area from  GRB 221009 A intersects the excluded area from the supernovae limits, at masses between  $\simeq {\rm a\, few}\,10$-$10^3\,\rm keV$, in the same mixing interval. However, GRB 221009 A excludes a large parameter of space above a mixing of $10^{-5}$. It is important to notice that the GRB 221009 A gives a better constraint than the supernova for a scalar with a mass of $m_S=10^3$ keV at a fixed mixing of $\sin\theta\simeq 10^{-8}$.

The time delay $\Delta t$ in arrival (or time of flight difference) to travel a distance $d$ between a massive particle of mass $m_S$ and a massless particle is given by the simple formula $\Delta t=\frac{m_S^2}{2 E_S^2} d$. We require that the time delay is smaller than the GRB duration of 2000s. We can see from this that the largest time delay is controlled by the higher mass scalars with lower total energy. To approximate the time delay, we consider the limiting mass for an average energy in the scalar spectrum of $\overline{E}_S$, which yields $m_S\simeq 4.5\times 10^3$ keV, well outside our allowed region in Fig.~\ref{fig:Limits1} which is cut-off at $10^3$ keV. Hence the allowed region is not affected by the time delay and complies easily with the 2000s requirement. If, however, additional information arises enabling the identification of the lowest photon energy in the range as low (or lower) than 0.5 TeV from $S$ decays, this would strengthen the bound to lower masses, although such an identification is unlikely given the large background from conventional photons in this region. Conversely, if a very high energy photon is identified with a long time delay event, this would indicate higher $m_S$. For massless particles in the final state produced by the decay of a massive boosted scalar, like in our scenario, the angular dispersion $\Theta$ in extremely relativistic limit in the lab frame is given in general by $\Theta=\frac{2m_S}{E_S}$. Taking the maximum mass of interest as a limiting scenario and the average scalar energy, we get $\Theta\simeq 10^{-7}$. Hence, as long as the GRB jet opening angle exceeds this tiny number, there is no additional suppression of the $\gamma$-rays from $S$ at the Earth.

\section{Conclusion}
\label{sec:conclusion}

In this work, we have considered  a generic scalar CP-even $S$ scalar corresponding to a CP-even singlet extension of the SM scalar sector as an explanation for GRB 221009 A.
We consider a scenario wherein the $S$ particles are produced in the GRB through hadronic scattering and then undergo the radiative loop-level decay  $S\rightarrow \gamma\gamma$ while propagating to Earth. If the photons are produced  without being nullified by the extragalactic background, it could explain the anomalously high observed $\gamma-$ray emission at 18 TeV. 

We calculated the flux of $\gamma$-rays from this process for different benchmarks  of the expected number of events on Earth. We found that the $\gamma$-ray flux produced increases at higher photon energies and approaches the unattenuated $\gamma$-ray flux. 

We have also computed bounds on the mixing-mass parameter space of the $S$ scalar for different fractions of the observed LHAASO events. We found that  at masses between  $\simeq {\rm a\, few}10$-$10^3\,\rm keV$, it overlaps with the excluded parameter space set by the SN1987A luminosity bounds. 
However, an upper bound exists in the parameter space of large mixing for all masses between $\simeq 10^1$-$10^3\,\rm keV$ that is not excluded by SN1987A or collider measurements. We also found that the strongest $S$ constraint for GRB 221009 A is for a scalar with a mass of $10^3$ keV and a mixing of $\simeq 10^{-8}$ for 50 events.

\color{black}
\section{Acknowledgements}
 S.B. would like to thank Andrea Caputo for helpful discussion. S.B. is supported by funding from the European Union’s Horizon 2020 research and innovation programme under grant agreement No.~101002846 (ERC CoG ``CosmoChart'') as well as support from the Initiative Physique des Infinis (IPI), a research training program of the Idex SUPER at Sorbonne Universit\'{e}.
 M.R.Q. is supported  by the JSPS KAKENHI Grant Number 20H01897.
Y.Z. is supported by the National Natural Science Foundation of China under grant No.\  12175039, the 2021 Jiangsu Shuangchuang (Mass Innovation and Entrepreneurship) Talent Program No.\ JSSCBS20210144, and the ``Fundamental Research Funds for the Central Universities.''
\bibliographystyle{apsrev4-1}
\bibliography{references.bib}

\end{document}